\documentclass[11pt,a4paper,oneside]{article}
\usepackage{epsfig,moriond}
\usepackage{amsmath,bm}

\newcommand{\as}{\alpha_{\rm s}}

\begin{document}
\vspace*{1.6cm}

\title{MATCHING PARTON SHOWERS TO NLO COMPUTATION}
\author{ZOLT\'AN NAGY}
\address{
Institute for Theoretical Physics, University of Z\"urich\\
Winterthurerstrasse 190, 
CH-8057 Z\"urich, Switzerland\\ 
}

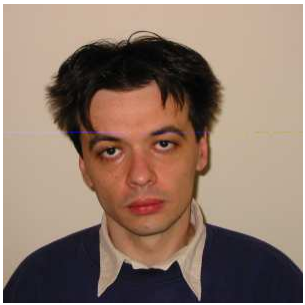
\maketitle\abstract{
In this review a new method is presented for attaching parton shower
algorithms to NLO partonic jet cross sections in electron-positron
annihilation. Our method is based on the Catani-Seymour dipole
subtraction method and also uses an adaptation of the matching scheme of
Catani, Krauss, Kuhn, and Webber. 
}


\section{Introduction}
\label{sec:Introduction}

One often uses perturbation theory to produce predictions for the results
of particle physics experiments in which the strong interaction is
involved. In order to get useable predictions one has to calculate at least at 
next-to-leading order to avoid large uncertainties those come from the unphysical scale dependences.  

Unfortunately, standard NLO programs have significant flaws. One flaw is 
that the final states consist just of a few partons, while in nature final states consist of many hadrons. A worse flaw is that the weights are often 
very large positive numbers or very large negative numbers. 

There is another class of calculational tools, the shower Monte Carlo
event generators, such as H{\small ERWIG} \cite{Herwig} and P{\small
YTHIA} \cite{Pythia}.  These have the significant
advantage that the objects in the final state consist of hadrons.
Furthermore, the weights are never large numbers. Finally, the programs 
have a lot of important structure of QCD built into them. For at least some cases like this, the shower Monte Carlo programs can provide a good approximation for the cross sections. The chief disadvantage of typical shower Monte Carlo event generators is that they are based on leading order
perturbation theory for the basic hard process and thus reproduce only
the first term in the perturbative expansion when applied to an infrared
safe observable.

It is possible to add the machinery of a shower Monte Carlo
event generator to a next-to-leading order program in such a way that the
complete program produces realistic final states made of hadrons, with
weights are not unbounded in size. One example is the program of Frixione, Nason, and Webber \cite{FrixioneWebberI}, which so far
has been applied to cases with massless incoming partons but not to
cases with massless final state partons at the Born level
of calculation. The other example is that of \cite{nloshowersI}, which concerns three-jet observables in
electron-positron annihilation and thus addresses massless final state
partons but not massless initial state partons.

In this paper, we want, most of all, to have an algorithm that can be used 
by NLO practitioners in a reasonably straightforward manner. For this reason, we
have based the algorithm on the dipole subtraction scheme of Catani and
Seymour \cite{CataniSeymour}.  It is quite widely used for NLO calculations (for
example in the programs N{\small LOJET++} \cite{Nagy} and M{\small CFM}
\cite{MCFM}).

We note that NLO calculations are generally limited to just one class of
observables -- for instance four-jet production but not at the same time
three-jet production. We would like to overcome this limitation. For this
reason we have adapted the $k_T$-jet matching scheme of Catani, Kuhn,
Krauss, and Webber \cite{CKKW} to the present circumstances.
We also seek be as independent as possible of the choice of any particular
shower Monte Carlo event generator. That is, we do not think that
practitioners of NLO calculations should need to do separate calculations
for each present and future shower Monte Carlo. 

The general idea of the algorithm that we present applies,
we believe, to lepton-lepton collisions, lepton-hadron collisions, and
hadron-hadron collisions. 

In the next section we have a very brief review of the algorithm. The precise definition and the details of the algorithm can be found in Ref.~\cite{Us}.

\section{Structure of the algorithm}
\label{sec:PartialCrossSectionswithShowers}

The cross section computed with parton showers will consist of
contributions from each available $m$,
\begin{equation}
\sigma^{\rm NLO+S} = \sum_{m=2}^{m_{\rm NLO}}
\left[
\sigma^{\rm B+S}_{m} + \sigma^{\rm R+S}_{m} + \sigma^{\rm V+S}_{m} 
\right]
+ \sum_{m=m_{\rm NLO}+1}^{m_{\rm max}} \sigma^{\rm B+S}_{m}
\;\;.
\label{sigmaNLOplusS}
\end{equation}
For the contributions at NLO level, there are three terms, which
correspond to Born, real emission, and virtual loop  contributions
with showers added (``$+S$''). For the remaining terms there is only a 
Born contribution. We will arrange that (for a suitably behaved
observable)
\begin{equation}
\sigma^{\rm B+S}_{m}  + \sigma^{\rm R+S}_{m} + \sigma^{\rm V+S}_{m}
=
\sigma^{\rm NLO}_{m} 
+{\cal O}(\alpha_{\rm s}^{B_m + 2})
+{\cal O}(1\, {\rm GeV}/\sqrt s)\;\;,
\label{showernet}
\end{equation}
that the NLO expansion of the partial shower cross sections gives the correct 
NLO partonc cross section ($\sigma^{\rm NLO}_{m}$) plus higher order and power corrections.

\subsection{Born term with showers}
\label{sec:BornwithShowers}

Our discussion begins in this subsection with $\sigma^{\rm B +
S}_m$. We define
\begin{equation}
\begin{split}
  \sigma^{B+S}_{m} = {}& \frac{1}{m!}  \sum_{\{f\}_m}
  \int\!d\varGamma(\{p\}_m)\, 
  \theta(d_{\rm ini}< d_m(\{p,f\}_m))\,W_{m}(\{p,f\}_m)
  \\
  &\times \sum_{l=1}^m \sum_{k\ne l}
  \big\langle {\cal M}(\{p,f\}_m)\big| 
   \int\!dY_l\
  \bm{E}_{l,k}(Y_l)\, \big|{\cal M}(\{p,f\}_m)\big\rangle
  I(\{p,f\}_m;l,k,Y_l) \;\;.
\label{sigmaBS}
\end{split}
\end{equation}
The first line contains integrals over Born level parton momenta 
($d\varGamma(\{p\}_m)$)and a
corresponding sum over parton flavors. The second line contains sums over
choices $l$ of the parton that splits and a spectator parton $k$ along
with an integral over the splitting variables\footnote{The splitting variables are the $y_{l}$ virtuality like variable, $z_{l}$ momentum fraction variable, $\phi_{l}$ azimuthal angle and the flavors of the emitted partons. The integral over these variables is 
\begin{equation*}
	\int_0^1 \frac{dy_l}{y_l}
	\int_0^1\!dz_l\,\int_0^{2\pi}\frac{d\phi_l}{2\pi}\,
	\frac{1}{2}\!\sum_{\hat f_{l,1},\hat f_{l,2}}\!
	\delta_{\hat f_{l,1} + \hat f_{l,2}}^{f_l}
	\equiv \int\! dY_l\;\; .
	\label{dYldef}
\end{equation*}
}
$Y_{l} = \{y_l, z_l,\phi_l,\hat f_{l,1},\hat f_{l,2}\}$ and a matrix element
of certain operators $\bm{E}_{l,k}$ acting on the Born amplitude
$|{\cal M}(\{p,f\}_m)\rangle$. The operators $\bm{E}_{l,k}$ together with the other factors in the formula describe the formation of showers from the Born
level partons. 

The integration over the momenta $\{p\}_m$ is restricted by a factor
$\theta(d_{\rm ini}< d_m(\{p,f\}_m))$.
Here $d_m(\{p,f\}_m)$ is defined by applying the $k_T$ jet finding
algorithm to the $m$ parton momenta.
Given an $n$-parton final state, we apply the recursive ``$k_T$'' jet
finding algorithm \cite{kTalgorithm} to the parton momenta $\{p, f\}_n$,
successively grouping the partons into jets. 
The algorthm  gives a sequence of jet resolution
parameters $d_J(\{p,f\}_n)$ at which two jets were joined, reducing $J$
jets to $J-1$ jets. Typically one has $d_n < d_{n-1} < \cdots < d_3$. 

There is also a factor $W_{m}(\{p,f\}_m)$, which is the product of
factors associated with the splitting history that matches the
found jet structure, following the method of Ref.~\cite{CKKW}. 

The function $I(\{p,f\}_m;l,k,Y_l)$ is the interface function to secondary shower. Its important property is that the secondary shower provides only perturbative and power correction, thus we have
\begin{equation}
I(\{p,f\}_m;l,k,Y_l) = F_{m+1}(\{\hat p\}_{m+1}) + {\cal O}(\as) 
+ {\cal O}(1\,{\rm GeV}/\sqrt{s})\;\;,
\end{equation}
where the function $F_{m+1}(\{\hat p\}_{m+1})$ is the measurement function of an infrared safe observable.

Now, we turn to the splitting function $\bm{E}_{l,k}$ in
Eq.~(\ref{sigmaBS}), which is an operator on the flavor and spin space of
parton $l$ in the vector $|{\cal M}\rangle$. This operator has the
following form,
\begin{equation}
\label{Elfinal}
\begin{split}
  \bm{E}_{l,k} = {}& \frac{\bm{T}_{l}\cdot
    \bm{T}_{k}}{-\bm{T}_{l}^2}\, 
 \int_{0}^{\infty}\! dr\,
   \delta(r - R_{l}(\{p,f\}_{m},y,z))\,
   \theta(\tilde d(\{p,f\}_m, l, y_l,z_l) < d_{\rm ini}) \,\frac{\alpha_{\rm s}(r)}{2\pi}
  \\ &\times
  \bm{S}_{l}(p_l,f_l,Y_l)
  \exp\!\biggl( -\int_{r}^{\infty}\!\! dr' \int_{0}^1
  \frac{dy'}{y'}\int_0^1\!\! dz'\, 
   \sum_{l'}
  \delta(r' - R_{l'}(\{p,f\}_{m}, y',z'))
  \\
  &\hskip 3.3 cm \times 
 \theta(\tilde d(\{p,f\}_m, l', y',z') < d_{\rm ini})\,
  \frac{\alpha_{\rm s}(r')}{2\pi}\,
  \big\langle\bm{S}(y',z',f_{l'})\big\rangle \biggr)\;\;.
\end{split}
\end{equation}
The parton splitting is organized according to an evolution parameter
$r$, which is defined to be proportional to the transverse momentum square,
$R_{l}(\{p,f\}_{m}, y,z) = s_l\, y z(1-z)$ and $s_l$ is a virtuality scale appropriate to parton $l$. The simplest choice would be $s_l = s$. .

With the use of function\footnote{
It is an approximation of the jet resolution variable,
\begin{equation*}
  \tilde d(\{p,f\}_m,l,y_l,z_l) = \frac{s_l}{s}\,
  y_{l} \min\left\{\frac{1-z_{l}}{z_{l}},
\frac{z_{l}}{1-z_{l}}\right\}\;\;.
\label{tildeddef}
\end{equation*}
}
$\tilde d(\{p,f\}_m, l, y_l,z_l)$, we
limit the splitting in $\bm{E}_{l}$ to be unresolvable at a scale $d$ that
is approximately $d_{\rm ini}\times 2p_l\cdot p_{k_l}/s_l$. With this cut the vetoing procedure is implemented in the first step of the shower ensuring the cancellation of the $d_{\rm ini}$ dependences al least at NLL level.

In each $\bm{E}_{l,k}$ operator, there is an operator on the parton color
space, ${\bm{T}_l\cdot \bm{T}_{k}}/[{-\bm{T}_l^2}]$, that is for the soft color connections. The splitting function $\bm{S}_{l}$ acts in the spin space of the emitter and depends on the splitting parameters  $Y_l$
for parton $l$ as well as on the momentum $p_l$. This functions are proportional to the dipole splitting functions.

The next factor, the Sudakov exponential, gives the probability that
{\it none} of the partons has split at a higher evolution scale. 
The factor $\langle \bm{S}(y_{l'},z_{l'},f_{l'})\rangle$ in the Sudakov 
exponent is the average over angle and flavors of $\bm{S}$ for parton $l'$.

\subsection{NLO corrections with shower}
\label{sec:NLOrealonShower}

We turn to the discussion of the NLO corrections. Let us start with
the real contribution. Define
\begin{equation}
  \label{sigmaRS}
  \begin{split}
    \sigma^{R+S}_{m} = {}& \frac{1}{(m+1)!}\sum_{\{\hat f\}_{m+1}}
    \int\!d\varGamma(\{\hat p\}_{m+1})\, 
     \tilde I\big(\{\hat p,\hat f\}_{m+1}\big)
    \\
    &\times\bigg\{ \big|{\cal M}(\{\hat p,\hat f\}_{m+1})\big|^{2}\,
    \theta\big(d_{m+1}(\{\hat p, \hat f\}_{m+1}) < d_{\rm ini}
    	< d_{m}(\{\hat p, \hat f\}_{m+1})\big)
    \\
    &\qquad -\sum_{\substack{{i,j}\\{\rm pairs}}} \sum_{k\neq i,j}
    {\cal D}_{ij,k} \,
    \theta\big(\tilde d(\{p,f\}_{m}^{ij,k}, \{l,y,z \}_{ij,k}) 
            < d_{\rm ini}< d_{m}(\{p,f\}_{m}^{ij,k})\big)
    \bigg\}\;\;.
  \end{split}
\end{equation}
The first term is the $m+1$ parton matrix element squared with the proper $m$-jet definition and the second
term is the sum of the dipole contributions to eliminate the infrared
singularities. The dipole function ${\cal D}_{ij,k}$ are based on the $m$-parton tree level color connected matrix elements and the correct definition can be taken from the paper by Catani and Seymour \cite{CataniSeymour}.

The definition of the virtual correction is 
\begin{equation}
\begin{split}
\sigma^{\rm V + S}_m
= {}& 
\frac{1}{m!}
\sum_{\{f\}_m} 
\int\!d\varGamma(\{p\}_m)\
 \theta\big( d_{\rm ini} < d_{m}(\{p,f\}_{m})\big)\,
\tilde I(\{p,f\}_m)\,
\\ &\times
\biggl\{
V(\{p,f\}_m)-
\frac{\alpha_{\rm s}(\mu_{\rm R})}{2\pi}\,
    W_{m}^{(1)}(\{p,f\}_m)\, \big|{\cal M}(\{p,f\}_m)\big|^{2}
    \biggr\}\;\;,
\end{split}
\label{sigmaVS}
\end{equation}
where the function $V(\{p,f\}_m)$ represents sum of the 1-loop matrix element and integrated subtraction term given in the second term of Eq.~\eqref{sigmaRS} over the phase space of the unresolved particle. The function  $W_{m}^{(1)}(\{p,f\}_m)$ is coefficient of the $\as$ term in the expansion of Sudakov reweighting factor $W_{m}(\{p,f\}_m)$.

\subsection{Secondary shower}

Consider, the function $I(\{p,f\}_m;l,k,Y_l)$ used for
$\sigma_m^{\rm B + S}$. In this term the $\bm{E}$ operator describe the emission of the hardest splitting, thus all the further splittings are constrained.   

All of the partons are allowed to split, and the one that does is
parton $l$ with aid of spectator $k$. The others did not split at an evolution variable above the value $r$. That is, parton $l'$, with the aid of
spectator $k'$, did not split with $r' > r$. Further evolution of these
partons should then be restricted to the range $r'<r$. This constraint for the transverse momentums is
\begin{equation}
|k_\perp^{\prime 2}| < 
\frac{2 p_{l'}\cdot p_{k'}}{s_{l'}}\ 
\frac{s_l}{2 p_l\cdot p_k}\
|k_\perp^{2}|\;\;.
\label{initialconditionA}
\end{equation}
A restriction like this can be imposed in the chosen shower Monte Carlo
program by using a veto algorithm, as described for instance in
Ref.~\cite{CKKW}.  A sensible choice for the $k'$ would be to let $k'$ be 
one of the final state partons to which parton $l'$ is color connected (at leading order in $1/N_{\rm c}$). 
For the splitting of one of the daughters of parton $l$, one may simply
impose $|k_\perp^{\prime 2}| <  |k_\perp^{2}|$.

Finally, in principle there should be a cut $d(p_i,p_j) < d_{\rm ini}$ imposed on further splittings. However, for most events passed to the Monte Carlo, 
$|k_\perp^{2}|$ will be much smaller than $d_{\rm ini} s$, so that this
cut is not really needed.

\section{Conclusion}

We have proposed an algorithm for adding showers to next-to-leading order
calculations for $e^+ + e^- \to N\ {\it jets}$. This algorithm is based
on the dipole subtraction scheme \cite{CataniSeymour} that is
widely used for next-to-leading order calculations. 
We also use the $k_T$-jet matching scheme of Ref.~\cite{CKKW} in order to
incorporate the possibility of calculating infrared safe $N$-jet cross
sections for different values of $N$ into the same computer program.


\end{document}